\documentclass[11pt]{article}
\usepackage{epsfig,here,hangcaption,rotating,fancyheadings,amssymb,amsmath}
\input epsf
\catcode`\@=11
\def\marginnote#1{}

\def\draftlabel#1{{\@bsphack\if@filesw {\let\thepage\relax
  \xdef\@gtempa{\write\@auxout{\string
    \newlabel{#1}{{\@currentlabel}{\thepage}}}}}\@gtempa
    \if@nobreak \ifvmode\nobreak\fi\fi\fi\@esphack}
     \gdef\@eqnlabel{#1}}
\def\@eqnlabel{}
\def\@vacuum{}
\def\draftmarginnote#1{\marginpar{\raggedright\scriptsize\tt#1}}
\def\draft{\oddsidemargin -.5truein
        \def\@oddfoot{\sl preliminary draft \hfil
        \rm\thepage\hfil\sl\today\quad\militarytime}
        \let\@evenfoot\@oddfoot \overfullrule 3pt
        \let\label=\draftlabel
        \let\marginnote=\draftmarginnote

\def\@eqnnum{(\theequation)\rlap{\kern\marginparsep\tt\@eqnlabel}%
\global\let\@eqnlabel\@vacuum}  }
\def\preprint{\twocolumn\sloppy\flushbottom\parindent 1em
        \leftmargini 2em\leftmarginv .5em\leftmarginvi .5em
        \oddsidemargin -.5in    \evensidemargin -.5in
        \columnsep 15mm \footheight 0pt
        \textwidth 250mmin      \topmargin  -.4in
        \headheight 12pt \topskip .4in
        \textheight 175mm
        \footskip 0pt

\def\@oddhead{\thepage\hfil\addtocounter{page}{1}\thepage}
        \let\@evenhead\@oddhead \def\@oddfoot{} \def\@evenfoot{}
}
\def\titlepage{\@restonecolfalse\if@twocolumn\@restonecoltrue\onecolumn
     \else \newpage \fi \thispagestyle{empty}\c@page\z@
        \def\thefootnote{\fnsymbol{footnote}} }
\def\endtitlepage{\if@restonecol\twocolumn \else  \fi
        \def\thefootnote{\arabic{footnote}}
        \setcounter{footnote}{0}}  
\catcode`@=12
\relax
\newcommand{\newc}{\newcommand}
\newc{\ie}{{\it i.e.}}
\newc{\eg}{{\it e.g.}}

\def\chioi{\tilde{\chi}^0_1}

\def\staui{\tilde{\tau}_1}
\def\slr{\tilde{l}_R}
\def\ohsq{\Omega_{\chi}h^2}
\def\tgbet{\mathrm{\tan}\beta}

\def\MT2{M_{T2}}

\def\m0{${0}$}

\def\tg{{\tilde g}}
\def\tq{{\tilde q}}

\def\tb{{\tilde b}}
\def\tchi{{\tilde\chi}}

\def\lsp{{\tilde\chi_1^0}}

\begin{document}
\topmargin-1.cm
%
\begin{titlepage}
\vspace*{-64pt}

\begin{flushright}
{SHEF-HEP/03-6,\\
February 2004}\\
\end{flushright}

\vspace{1.8cm}

\begin{center}

\large{{\bf Constraining SUSY Dark Matter with the ATLAS Detector at
the LHC}}\\
\vspace*{1.3cm}
\large{G. Polesello}\\
{\it\small INFN, Sezione di Pavia, Via Bassi 6, I-27100 Pavia,
Italy.}\\
\vspace*{0.4cm}
\large{D.R. Tovey}\\
{\it\small Department of Physics and Astronomy, University of
Sheffield, \\ Hounsfield Road, Sheffield S3 7RH, UK.}\\

\end{center}

\bigskip
\begin{abstract}
In the event that R-Parity conserving supersymmetry (SUSY) is
discovered at the LHC, a key issue which will need to be addressed
will be the consistency of that signal with astrophysical and
non-accelerator constraints on SUSY Dark Matter. This issue is studied
for the SPS1a mSUGRA benchmark model by using measurements of
end-points and thresholds in the invariant mass spectra of various
combinations of leptons and jets in ATLAS to constrain the model
parameters. These constraints are then used to assess the statistical
accuracy with which quantities such as the Dark Matter relic density
and direct detection cross-section can be measured. Systematic effects
arising from the use of different mSUGRA RGE codes are also
estimated. Results indicate that for SPS1a a statistical(systematic)
precision on the relic abundance $\sim$ 2.8 \% (3 \%) can be
obtained given 300 fb$^{-1}$ of data.
\end{abstract}





\vspace{1cm}

\vspace{-0.5 cm}


\end{titlepage}

\setcounter{footnote}{0}
\setcounter{page}{0}
\newpage

\section{Introduction}
The complementarity of LHC SUSY measurements and direct and indirect
searches for neutralino Dark Matter is an important topic to study
given the increasingly strong astrophysical evidence for cold dark
matter in the universe
\cite{Bennett:2003bz,Spergel:2003cb,Peiris:2003ff}. Assuming that
R-Parity conserving SUSY is discovered at the LHC, an interesting
question will arise regarding the compatibility of that signal with
existing relic density constraints (e.g. $0.094<\ohsq<0.129$ at
$2\sigma$ from WMAP data
\cite{Bennett:2003bz,Spergel:2003cb,Peiris:2003ff}), and the
implications it has for terrestrial Dark Matter searches.

In this paper we address these issues within the context of the
minimal supergravity (mSUGRA) class of SUSY models incorporating
gravity-mediated SUSY breaking \cite{msug}. A great advantage of
mSUGRA models when studying SUSY phenomenology is that they can be
described with only five independent paramters, namely the common
scalar mass at the GUT scale ($m_0$), the common gaugino mass
($m_{1/2}$), the common trilinear coupling parameter $A_0$, the value
of the ratio of the SUSY higgs vacuum expectation values
($\tan{\beta}$) and the sign of the higgsino mass parameter
(sgn($\mu$)). We choose to study one particular model referred to as
`SPS1a' ($m_0=100$ GeV, $m_{1/2}=250$ GeV, $A_0=-100$ GeV,
$\tan{\beta}=10$, $\mu>0$), which was defined in
Ref.~\cite{Allanach:2002nj}. This model lies in the `bulk' region of
the $m_0-m_{1/2}$ mSUGRA plane where the relic density is reduced to a
small value by $\chioi$ annihilation to leptons or neutrinos via
t-channel slepton, stau and sneutrino exchange. This model point was
defined prior to the latest WMAP data and consequently possesses a
relic density rather larger than that now favoured, however the
analysis described in this paper remains valid and provides an
interesting case study of the techniques which can be used when bulk
region SUSY models are chosen by Nature.

In Section 2 we describe the measurement of mSUGRA parameters at point
SPS1a using fits to measurements of end-points and thresholds in the
invariant mass spectra of various combinations of leptons and jets
observed in ATLAS \cite{Armstrong:1994it}. In Section 3 we discuss the
use of these parameter measurements to calculate both the $\chioi$
relic density and quantities of relevance to direct and indirect
search Dark Matter experiments. Finally in Section 4 we review the
results and discuss the circumstances under which the measurements can
be considered to be model-independent.

\section{Extraction of mSUGRA Parameters}
The development of techniques for measuring parameters characterising
SUSY models has been the subject of much investigation in the last few
years, as documented in Ref.~\cite{Armstrong:1994it,Bachacou:2000zb,
Allanach:2000kt}, and is still a very active field of investigation.

The basic issue is that the presence of two invisible particles in the
final state renders the direct measurement of sparticle masses through
the detection of invariant mass peaks impossible. Alternative
techniques have therefore been developed, based on the exclusive
identification of long cascades of two body-decays. It was
demonstrated \cite{Bachacou:2000zb} that if a sufficiently long chain
can be identified (at least three successive two-body decays), the
thresholds and end-points of the various possible invariant mass
combinations among the identified products can be used to achieve a
model-independent measurement of the masses of the involved
sparticles. Once the masses of the lighter sparticles are obtained
with this procedure, in particular the mass of the Lightest
Supersymmetric Particle (LSP), additional sparticle masses can be
measured through the identification of other shorter decay
chains. This program has been carried out recently for SPS1a
\cite{LHCLC}, resulting in a number of measurements of observables
which are related to the masses of the sparticles by known algebraic
relations (Table~\ref{tab:summes}). The meaning of the different
observables, and their expression in terms of sparticle masses is
given in Ref.~\cite{Allanach:2000kt}.
\begin{table}[htb]
\begin{center}
\vskip 0.2cm
\begin{tabular}{|l|c|c|c|c|}
\hline
& & \multicolumn{3}{c|}{Errors}\\
Variable & Value (GeV) & Stat+Sys (GeV) & Scale (GeV) & Total \\
\hline
\hline
$m_{\ell\ell}^{max}$            &      83.37   &   0.03  &   0.08  &    0.09\\
$m_{\ell\ell q}^{max}$          &     457.55   &   1.4  &    4.6  &    4.8 \\
$m_{\ell q}^{low}$              &     321.28   &   0.9  &    3.2  &    3.3\\
$m_{\ell q}^{high}$             &     400.63   &   1.0  &    4.0  &    4.1\\
$m_{\ell\ell q}^{min}$          &     220.81   &   1.6  &    2.2  &    2.7\\
$m_{\ell\ell b}^{min}$          &     199.48   &   3.6  &    2.0  &    4.2\\
$m(\ell_L)-m(\lsp)$             &     109.18   &   1.5  &    0.1  &    1.5\\
$m_{\ell\ell}^{max}(\tchi^0_4)$ &     279.07   &   2.3  &    0.3  &    2.3\\
$m_{\tau\tau}^{max}$            &      86.03   &   5.0  &    0.9  &    5.1\\
$m(\tg)-0.99\times m(\lsp)$     &     517.22   &   2.3  &    5.2  &    5.7\\
$m(\tq_R)-m(\lsp)$              &     452.62   &  10.0  &    4.5  &   11.0\\
$m(\tg)-m(\tb_1)$               &      96.98   &   1.5  &    1.0   &    1.8\\
$m(\tg)-m(\tb_2)$               &      72.75   &   2.5  &    0.7   &    2.6\\
\hline
\end{tabular}
\caption{\label{tab:summes} {\it Summary table of the SUSY
measurements which can be performed at the LHC with the ATLAS
detector. The central values are calculated for
ISAJET 7.69. The statistical errors are given for the integrated
luminosity of 300~$fb^{-1}$. The uncertainty in the energy scale is
taken to result in an error of 1\% for measurements including jets,
and of 0.1\% for purely leptonic mesurements. The values quoted here
differ somewhat from those quoted in Ref.~\cite{LHCLC} due to the use
of an earlier version of ISASUGRA in that work.}}
\end{center}
\end{table}
An additional measurement used in this analysis is the mass of the
light higgs boson from the decay \mbox{$h\rightarrow\gamma\gamma$}
which for 100~fb$^{-1}$, and for $m_h=114$~GeV, has a statistical
uncertainty of $\sim0.5$~GeV \cite{hohlfeld}.

Two approaches are possible in order to extract model constraints from
these measurements:
\begin{itemize}
\item
Solve the system of measurements for the masses of the involved
sparticles, and fit the model parameters to them;
\item
Perform a direct fit of the parameters of the model to the measured
observables.
\end{itemize}
As discussed in detail in \cite{Bachacou:2000zb, Allanach:2000kt}, the
mass values resulting from solving the experimental constraints are
strongly correlated. This is due to the fact that the measured
constraints are typically expressed as differences of masses.
Therefore the first approach, using uncorrelated mass errors, as is
e.g. done in Ref.~\cite{battaglia03} will result in a very inaccurate
estimate of the constraints on the model.

We rely therefore in this work on the direct fit of the observable to
the mSUGRA model as embodied in a RGE evolution code.  A detailed study
of the theoretical uncertainties based on the study of different
implementations of the RGE running of the SUSY masses from the GUT
scale to the TeV scale is given in Ref.~\cite{sabine}.  Addressing
this theoretical uncertainty is outside the scope of this work. We
limit ourselves to perform the fit on two different models, namely
{\tt ISASUGRA v.7.69}, and {\tt SUSPECT v.2.102}, with the aim of
studying the impact of the use of different SUSY spectrum generators
on the determination of the Dark Matter characteristics.

Two different types of uncertainties are quoted in Table
\ref{tab:summes}: the combined statistical and systematic
uncertainties estimated for each measurement, and general errors on
the scales of lepton and hadron energy measurement, which affect all
the measured quantities in the same way. Since in many cases the scale
uncertainties are dominant it is necessary to take into account the
correlations between the different measurements when extracting the
constraints on the model parameters. A technique often found in the
literature (see e.g. Ref.~\cite{sabine}) to estimate the error in
parameter determination may be described as follows:
\begin{enumerate}
\item
The central value $d(i)$ and measurement error $\sigma(i)$ is obtained
for each of the measured quantities, where $i$ runs over the number
$N_d$ of available measurements
\item
A scan is performed over a grid of SUSY models, with the nominal value
$dm(i,j)$ of each of the measurements $i$ being calculated for each
model $j$.
\item
The total $\chi^2$ is calculated for each point $j$ on the grid using
$$
\chi^2_j=\sum_i{\frac{d^2(i)^2-dm^2(i,j)}{\sigma^2(i)}}
$$
\item
The uncertainty in the hyperspace of SUSY parameters is evaluated
using the properties of the $\chi^2$ function.
\end{enumerate}

This approach relies on the assumption that the calculated $\chi^2$
function is correctly normalised, and the uncertainties in the
measurements are independent, however we know that this last
assumption is at best very approximate. We therefore use a Monte Carlo
technique relying on the generation of simulated experiments sampling
the probability density functions of the measured observables. In
frequentist statistics, confidence bands describe the probability that
an experiment in a set of identical experiments yields a given value
for the measured quantities.

\par We proceed in the following way:
\begin{enumerate}
\item
An `experiment' is defined as a set of 14 measurements, each of which
is generated by picking a value from a gaussian distribution with mean
given by the central value given in Table~\ref{tab:summes}. The
correlation due to energy scale is taken into account by using a
second gaussian distribution for the energy scale, and using the same
random number for all the measurements sharing the same scale.
\item
For each experiment, the point in mSUGRA parameter space minimising
the $\chi^2$ is calculated.
\end{enumerate}

\begin{figure}[htb]
\begin{center}
\epsfig{file=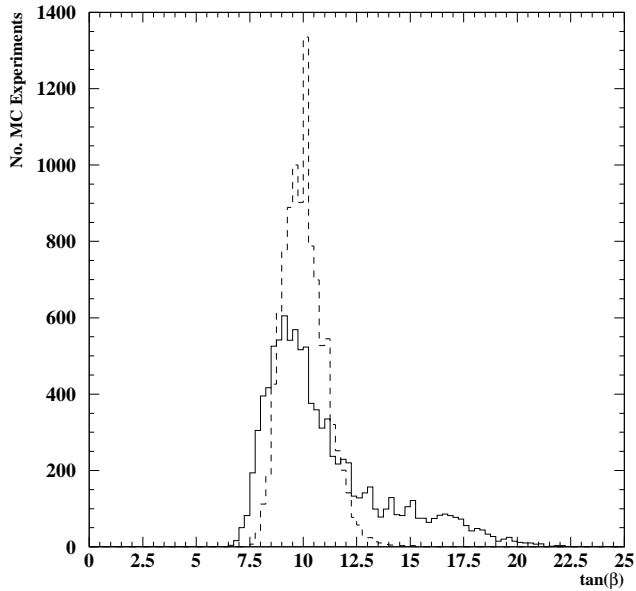,height=3.5in}
\caption{\label{fig0} {\it Distributions of fitted $\tgbet$ values
obtained using {\tt ISASUGRA v.7.69} assuming 100 fb$^{-1}$ of data
(full histogram) and 300 fb$^{-1}$ of data (dashed histogram). The
former distribution displays a significant tail to large values of
$\tgbet$.}}
\end{center}
\end{figure}

We obtain as a result of this calculation a set of mSUGRA models, each
of which is the ``best'' estimate for a Monte Carlo experiment of the
mSUGRA model generating the observed measurement pattern. For each of
these models the properties of the LSP Dark Matter candidate can then
be calculated and the spread of obtained results interpreted as the
level of precision with which these properties can be measured by the
LHC. For 300 fb$^{-1}$ of data the distributions of fitted mSUGRA
parameter values are all approximately gaussian in form with widths
$\sim$ 2\% ($m_0$), 0.6\% ($m_{1/2}$), 9\% ($\tgbet$) and 16\% ($A_0$)
respectively. For 100 fb$^{-1}$ of data the main differences are that
the width of the $A_0$ distribution increases to 28\% due to the
poorer precision on the $\tilde\chi^0_4$ mass and (more importantly)
that the $\tgbet$ distribution acquires a significant tail to large
values of $\tgbet$ (Fig.~\ref{fig0}) due to the fact that for 100
fb$^{-1}$ the masses of the two sbottom squarks can not be measured
separately, and are therefore not included in the fit. This tail has a
significant impact on the precision with which dark matter properties
can be calculated. Therefore most results quoted below assume 300
fb$^{-1}$ of data.

\section{Calculation of Dark Matter Characteristics} 
Best fit values of $m_0$, $m_{1/2}$, $A_0$ and $\tan(\beta)$ obtained
from each fit to the 14 sampled measurements were passed to dedicated
codes in order to calculate the implied characteristics of the
$\chioi$ Dark Matter candidate. As a result of the generally poor best
fit $\chi^2$ values obtained from the $\mu<0$ fits, the correct
$\mu>0$ assumption was made throughout. The input top mass values used
for each model were randomly sampled from a gaussian distribution of
mean 175 GeV and (conservative) $\sigma = 2$ GeV, to represent the
uncertainty arising from ATLAS measurements \cite{atltdr-top}.

\begin{figure}[htb]
\begin{center}
\epsfig{file=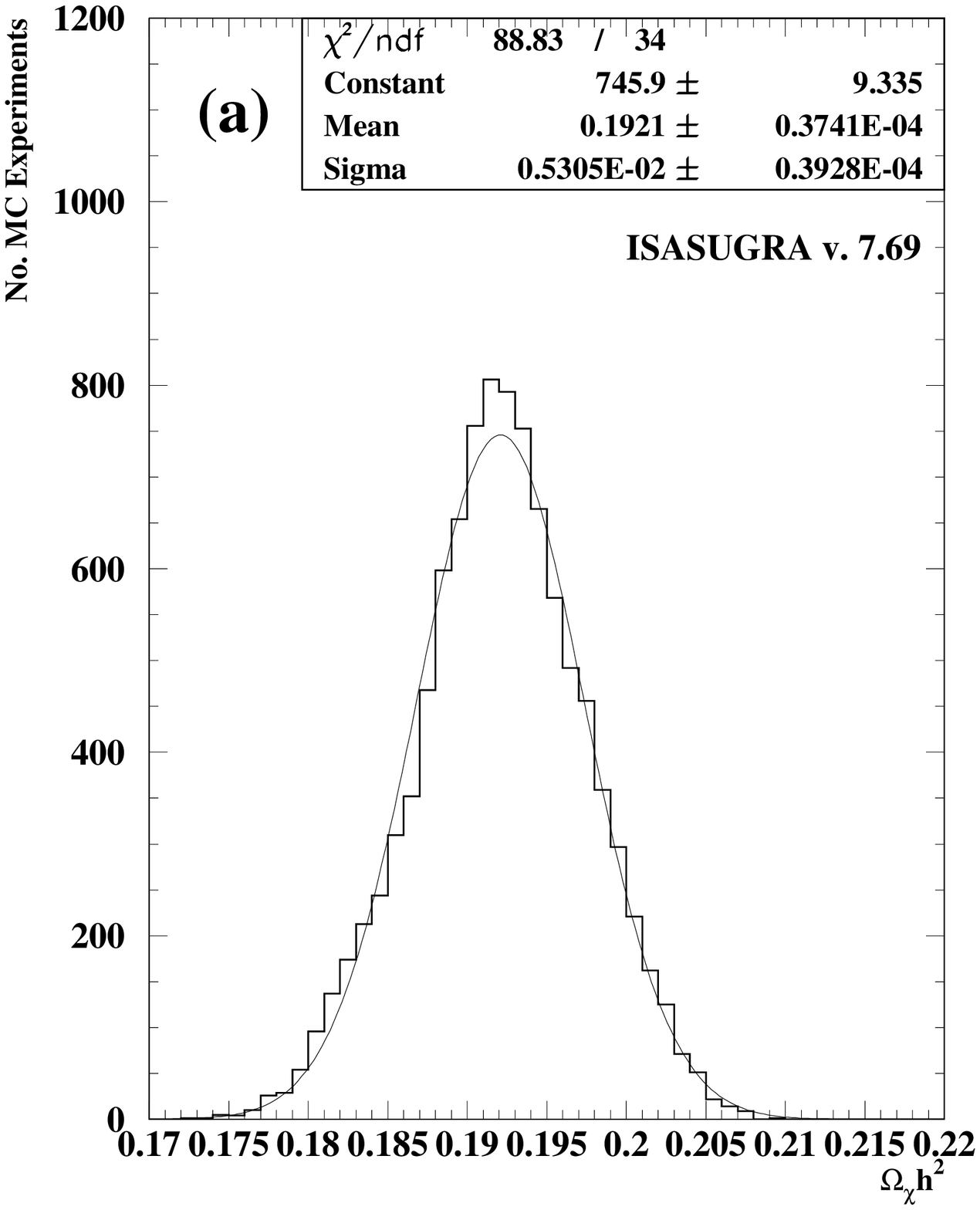,height=3.5in}
\epsfig{file=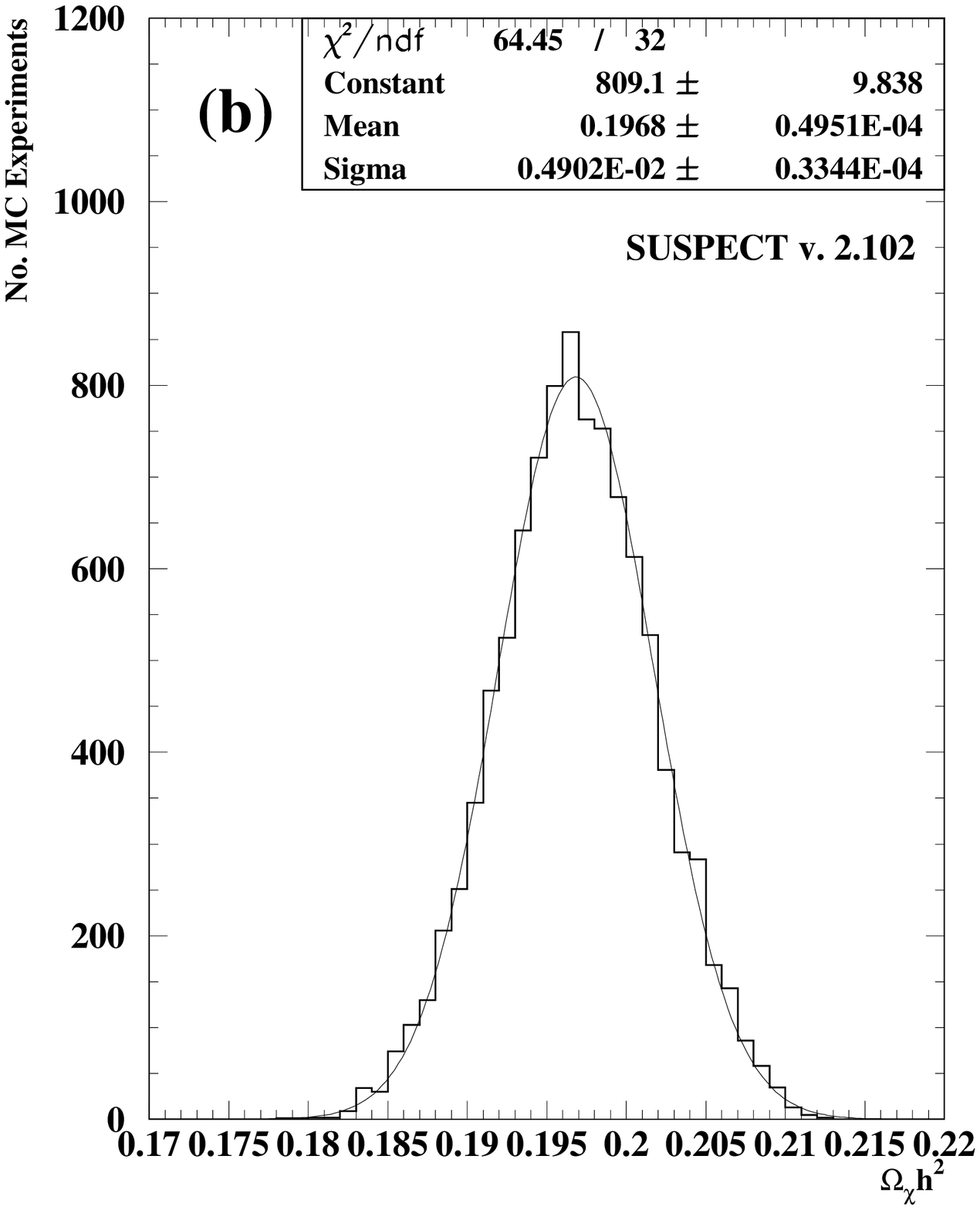,height=3.5in}
\caption{\label{fig1} {\it Values for $\Omega_{\chi} h^2$ calculated
from mSUGRA fits to the SPS1a invariant mass spectrum end-points
described in the text. The distribution in Figure (a) was calculated
by using results from {\tt ISASUGRA v.7.69} fits as input to {\tt
MICROMEGAS v.1.1.1} interfaced to {\tt ISASUGRA v.7.69}. The
distribution in Figure (b) was calculated by using results from {\tt
SUSPECT v.2.102} fits as input to {\tt MICROMEGAS v.1.1.1} interfaced
to {\tt SUSPECT v.2.102}.}}
\end{center}
\end{figure}

The $\chioi$ relic density $\ohsq$ was calculated using {\tt
MICROMEGAS v.1.1.1} \cite{Belanger:2001fz} interfaced to either the
{\tt ISASUGRA v.7.69} \cite{isa} or {\tt SUSPECT v.2.102}
\cite{Djouadi:2002ze} RGE codes. {\tt MICROMEGAS} incorporates exact
tree-level calculations of all annihilation and co-annihilation
processes together with one- and two-loop corrections to the Higgs
width and mass. {\tt MICROMEGAS} is hence well suited to this task at
SPS1a where stau annihilation processes dominate the relic density
calculation. Nevertheless an estimated 2 \% systematic error must be
assigned to the relic density due to uncertainties in solving the
Boltzmann equation \cite{gb}. An additional systematic uncertainty
$\lesssim$ 1 \% is estimated from higher order corrections to the
calculated cross-sections. Potential systematic uncertainties arising
from differences in the RGE codes were estimated from the mean
separation between the predicted $\ohsq$ distributions obtained from
the two codes.

Characteristics of potential Dark Matter detection signatures implied
by the ATLAS measurements were assessed using {\tt DarkSUSY v.3.14.02}
\cite{Gondolo:2000ee,Gondolo:2002tz}, again interfaced to the {\tt
ISASUGRA v.7.69} or {\tt SUSPECT v.2.102} RGE codes. {\tt DarkSUSY
v.3.14.02} is a relatively mature code which is in general unsuitable
for detailed relic density calculations due to its lack of an
implementation of e.g. stau co-annihilation processes \footnote{A new
version of {\tt DarkSUSY} incorporating co-annihilation processes is
under development \cite{Edsjo:2003us}.}. Nevertheless it is one of
only a few publicly available codes which can calculate quantities
relevant to direct and indirect detection. Three such quantities were
chosen for study in this paper, namely the spin-independent
$\chioi$-nucleon scattering cross-section $\sigma_{si}$ (equivalent to
the scalar scattering cross-section for Majorana neutralinos)
governing the rate of elastic nuclear recoils observed in direct
search experiments \cite{Goodman:1984dc}, the flux $\phi_{sun}$ of
high energy neutrinos generated by annihilation of relic $\chioi$
trapped in the solar core, and the flux $\phi_{earth}$ of neutrinos
generated by similar processes occurring within the earth
\cite{Jungman:1995df}. The last two quantities were calculated by
integrating over all neutrino energies for neutrinos emerging within a
half-aperture angle of 30$^o$ from the centre of sun/earth.

In all of the above cases care was taken to ensure that when using the
{\tt ISASUGRA} RGE code interfaced to a given Dark Matter code, input
best fit mSUGRA parameters were obtained using {\tt ISASUGRA}
fits. Similarly all results obtained using {\tt SUSPECT} made use of
{\tt SUSPECT}-derived input parameters.

\section{Results}

The distribution of $\ohsq$ values obtained from the best-fit mSUGRA
models by {\tt MICROMEGAS} interfaced to {\tt ISASUGRA} or {\tt
SUSPECT} are shown in Fig.~\ref{fig1}. As with all the other
quantities investigated here the results obtained with {\tt SUSPECT}
give a somewhat smaller statistical scatter ($\sim$ 2.5\%) compared
with those obtained with {\tt ISASUGRA} ($\sim$ 2.8\%). The overall
systematic uncertainty arising from the RGE codes is estimated to be
$\sim$ 2.4\%, giving an overall systematic uncertainty (adding the
Boltzmann contribution in quadrature) $\sim$ 3\%. In both cases the
distributions are reasonably gaussian.

\begin{figure}[htb]
\begin{center}
\epsfig{file=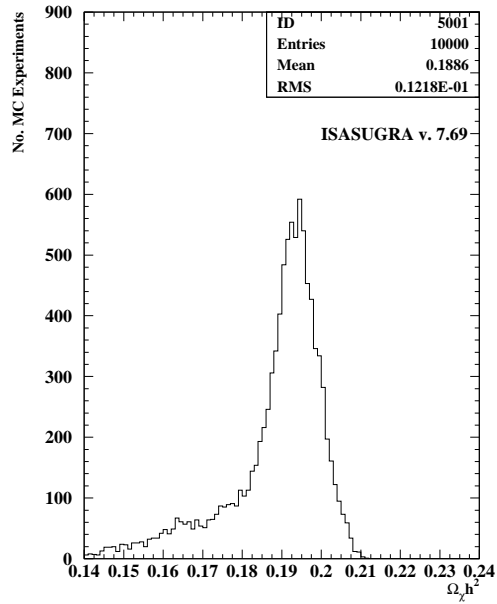,height=3.5in}
\caption{\label{fig1a} {\it Values for $\Omega_{\chi} h^2$ calculated
with {\tt ISASUGRA} and {\tt MICROMEGAS} in the same manner as for
Fig.~\ref{fig1}(a), except using mSUGRA parameter distributions
expected for 100 fb$^{-1}$ of data.}}
\end{center}
\end{figure}

It should be noted that with 100 fb$^{-1}$ of data the significant
tail in the $\tgbet$ distribution arising from the lack of a
measurement of the $\tilde{b}_1$ mass leads to significant
non-gaussianity in the prediction for $\ohsq$ (Fig.~\ref{fig1a}). The
same behaviour is also observed for the other dark matter quantities
considered below.

\begin{figure}[htb]
\begin{center}
\epsfig{file=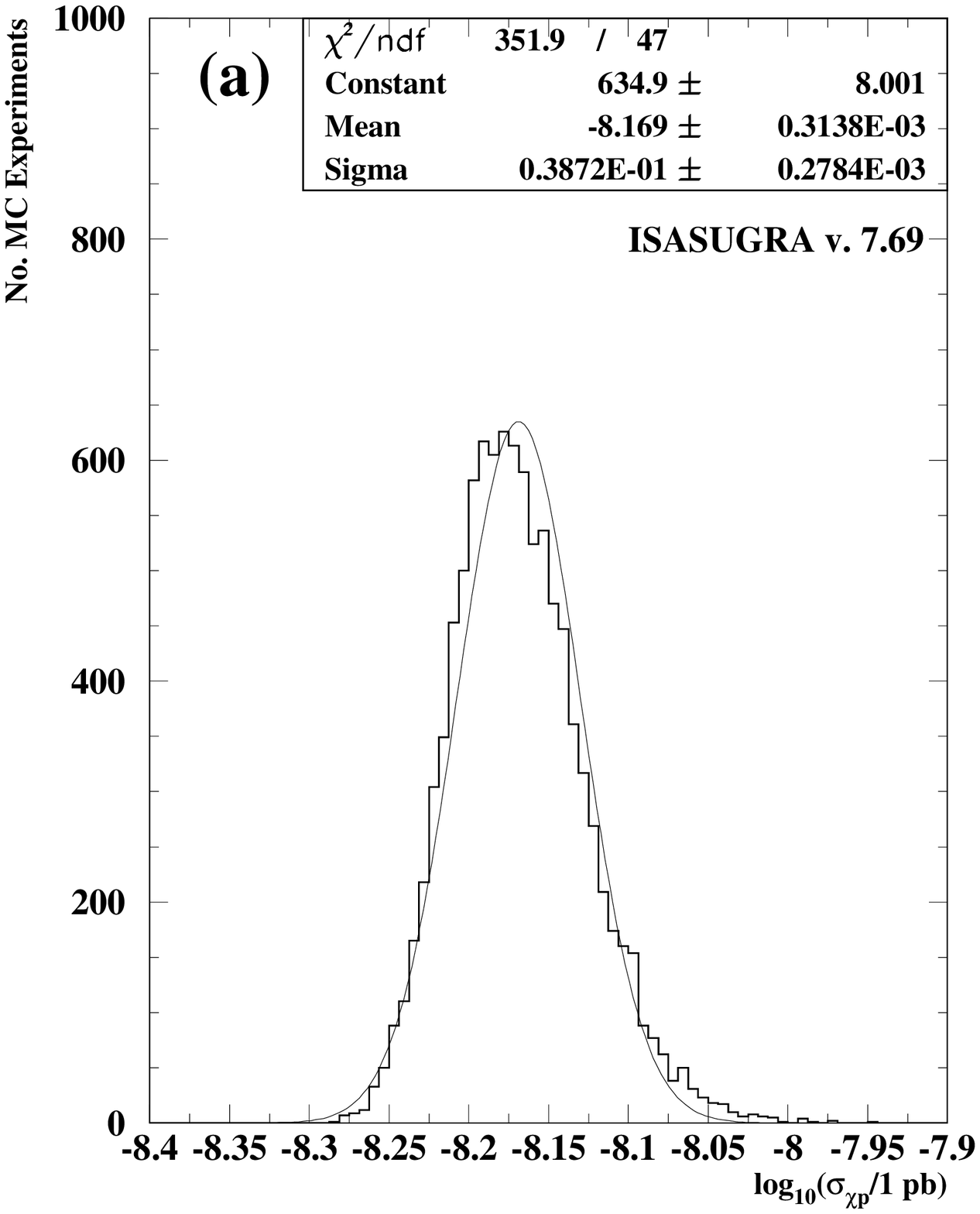,height=3.5in}
\epsfig{file=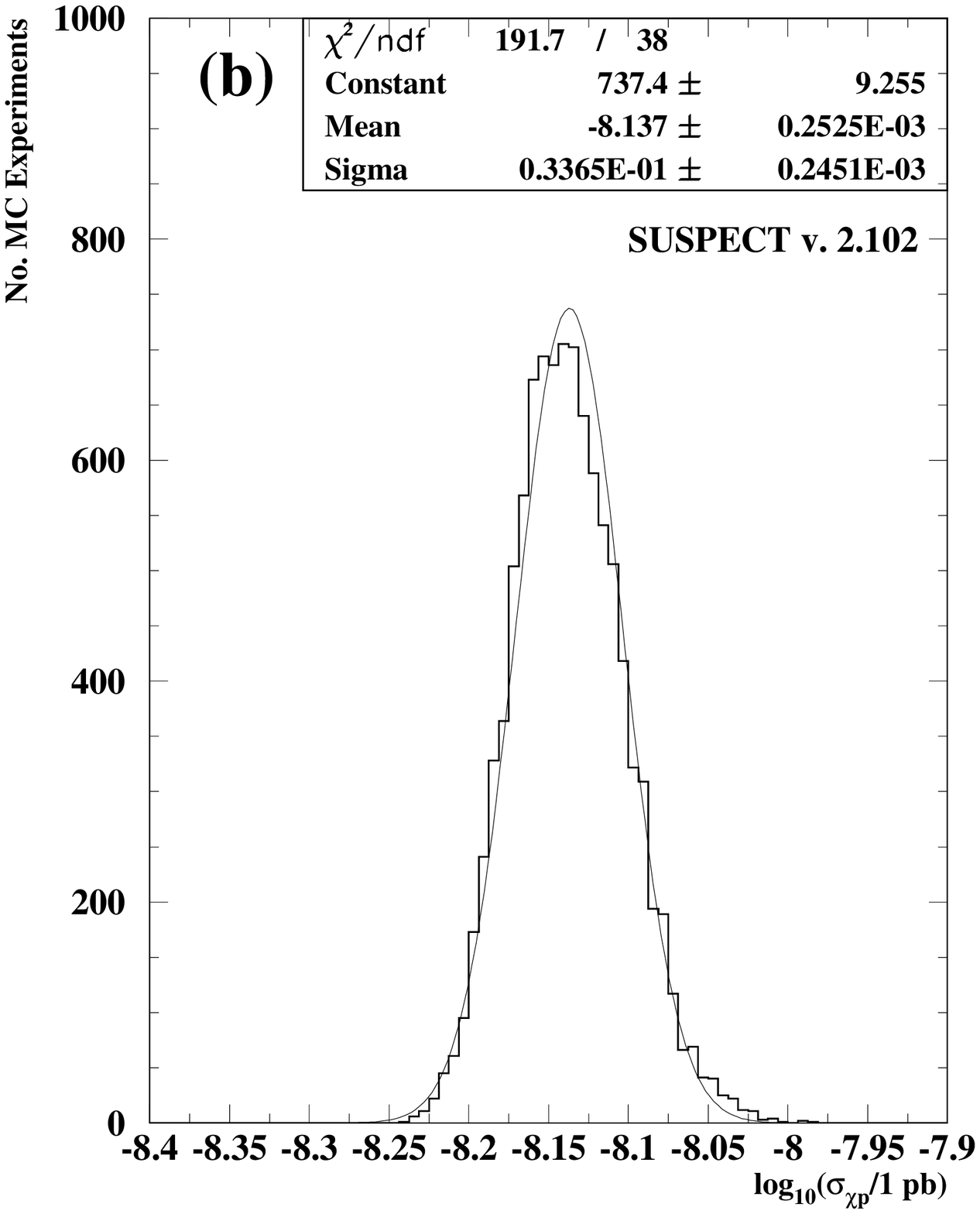,height=3.5in}
\caption{\label{fig2} {\it Values for the spin-independent
$\chioi$-nucleon elastic scattering cross-section calculated from
mSUGRA fits to the SPS1a invariant mass spectrum end-points described
in the text. The distribution in Figure (a) was calculated by using
results from {\tt ISASUGRA v.7.69} fits as input to {\tt DarkSUSY
v.3.14.02} interfaced to {\tt ISASUGRA} v.7.69. The distribution in
Figure (b) was calculated by using results from {\tt SUSPECT v.2.102}
fits as input to {\tt DarkSUSY v.3.14.02} interfaced to {\tt SUSPECT
v.2.102}.}}
\end{center}
\end{figure}
The distribution of log$_{10}\sigma_{si}$ values ($\sigma_{si}$
measured in pb) obtained from the best-fit mSUGRA models by {\tt
DarkSUSY} interfaced to {\tt ISASUGRA} or {\tt SUSPECT} are shown in
Fig.~\ref{fig2}. This time the statistical errors are even smaller
($\sim$0.5\%) with an RGE systematic error estimated to be \mbox{$\sim$
0.4}\%. These errors are expected to be negligible compared with the
intrinsic estimated factor 2 systematic uncertainty arising from
uncertainties in the strange quark content of the nucleon and
approximations used when calculating higher order gluon loop
contributions to the cross-section \cite{pgpriv}. Furthermore any
experimental measurement of $\sigma_{si}$ is expected to be subject to
considerable systematic uncertainty arising from lack of knowledge of
the local Dark Matter density and velocity distribution. Consequently
such measurements are not expected to agree with ATLAS predictions at
the high level of accuracy quoted here.

\begin{figure}[htb]
\begin{center}
\epsfig{file=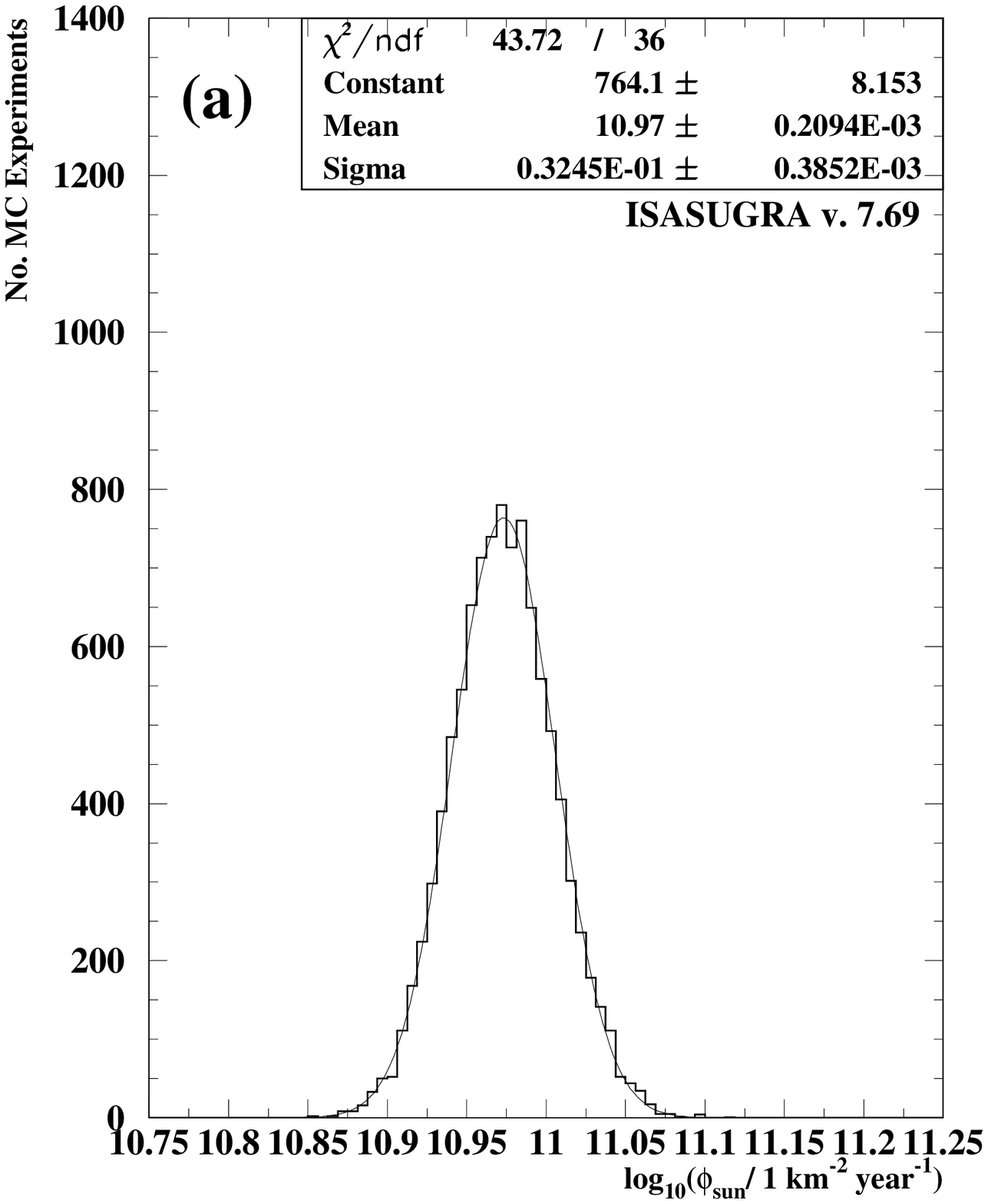,height=3.5in}
\epsfig{file=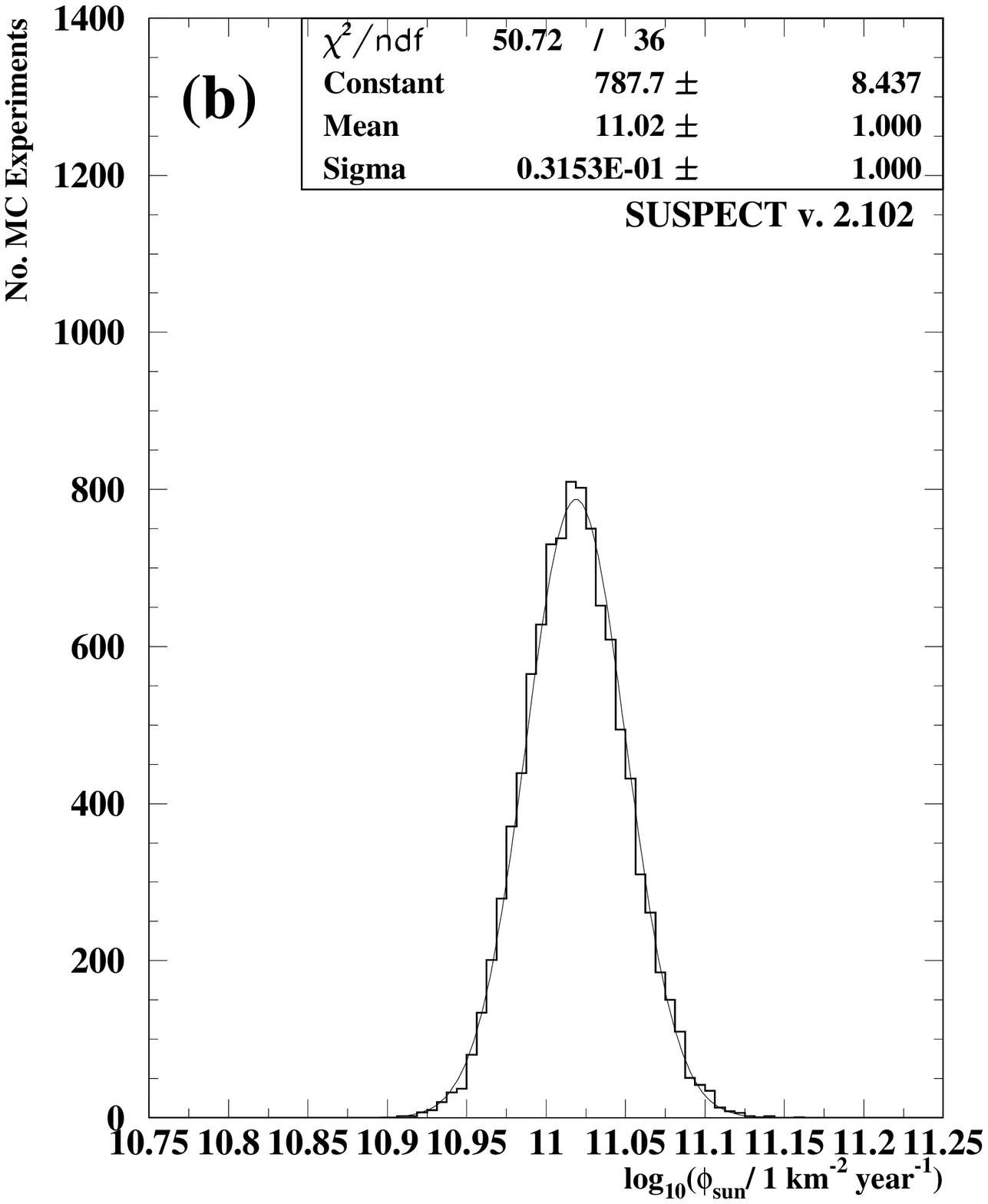,height=3.5in}
\caption{\label{fig3} {\it Values for the flux of high energy
neutrinos from $\chioi$-$\chioi$ annihilation in the sun calculated
from mSUGRA fits to the SPS1a invariant mass spectrum end-points
described in the text. The distribution in Figure (a) was calculated
by using results from {\tt ISASUGRA v.7.69} fits as input to {\tt
DarkSUSY v.3.14.02} interfaced to {\tt ISASUGRA} v.7.69. The
distribution in Figure (b) was calculated by using results from {\tt
SUSPECT v.2.102} fits as input to {\tt DarkSUSY v.3.14.02} interfaced
to {\tt SUSPECT v.2.102}.}}
\end{center}
\end{figure}

\begin{figure}[htb]
\begin{center}
\epsfig{file=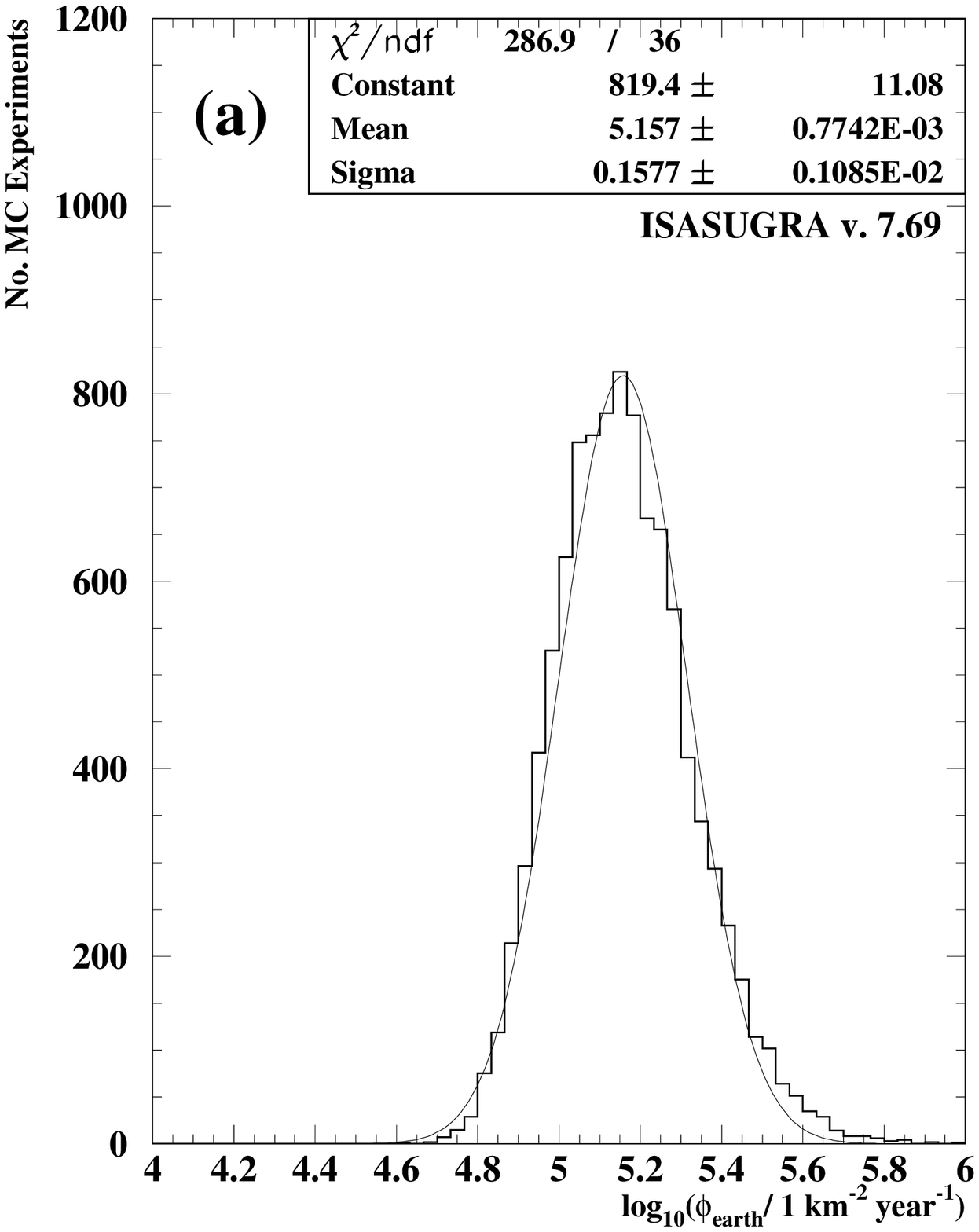,height=3.5in}
\epsfig{file=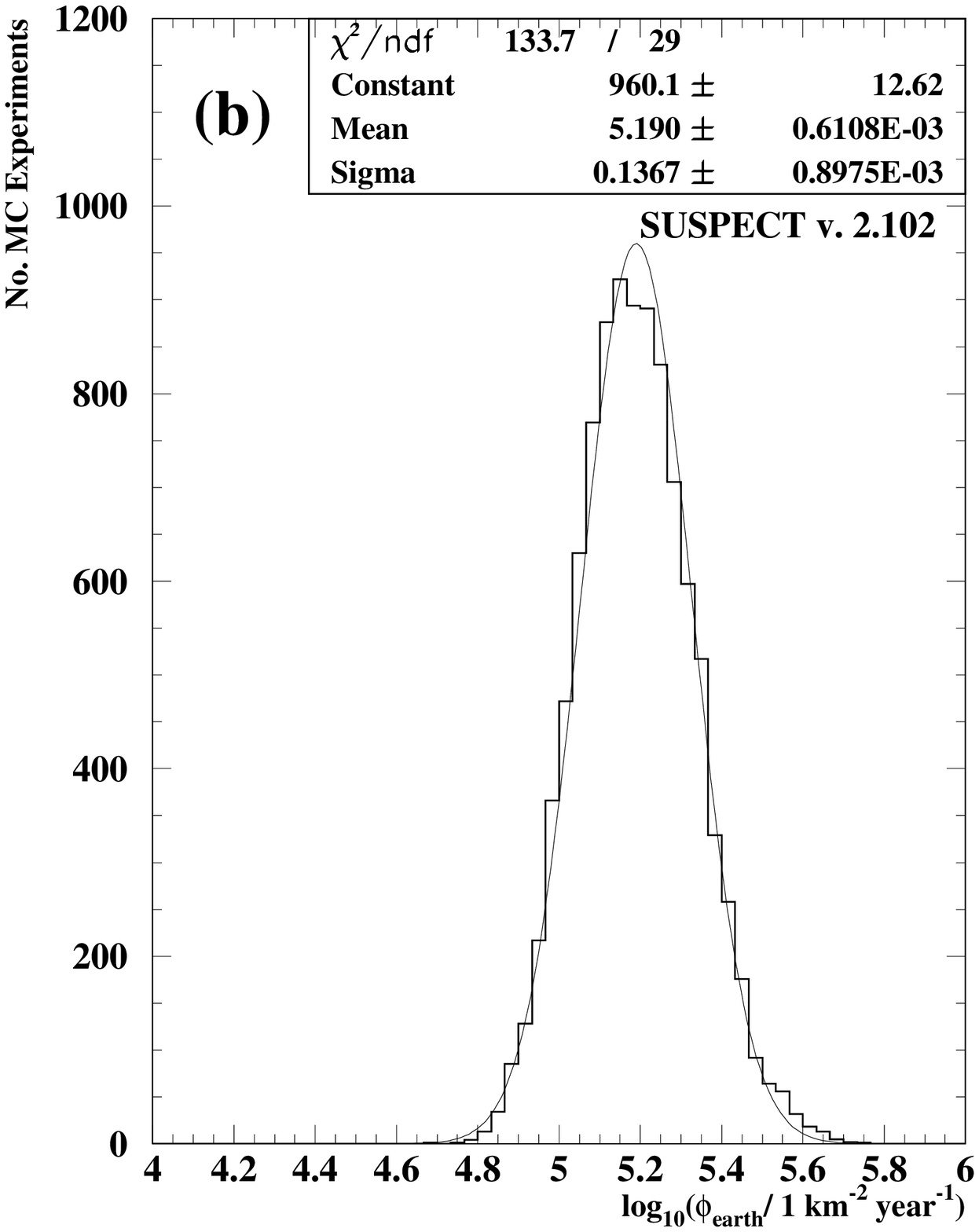,height=3.5in}
\caption{\label{fig4} {\it Values for the flux of high energy
neutrinos from $\chioi$-$\chioi$ annihilation in the earth calculated
from mSUGRA fits to the SPS1a invariant mass spectrum end-points
described in the text. The distribution in Figure (a) was calculated
by using results from {\tt ISASUGRA v.7.69} fits as input to {\tt
DarkSUSY v.3.14.02} interfaced to {\tt ISASUGRA} v.7.69. The
distribution in Figure (b) was calculated by using results from {\tt
SUSPECT v.2.102} fits as input to {\tt DarkSUSY v.3.14.02} interfaced
to {\tt SUSPECT v.2.102}.}}
\end{center}
\end{figure}
The distribution of log$_{10}\phi_{sun}$ and log$_{10}\phi_{earth}$
values ($\phi$ measured in km$^{-2}$ year$^{-1}$) obtained from the
best-fit mSUGRA models by {\tt DarkSUSY} interfaced to {\tt ISASUGRA}
or {\tt SUSPECT} are shown in Figs.~\ref{fig3}
and~\ref{fig4}. Statistical errors $\sim$ 0.3\% (3\%) were obtained
for the flux of neutrinos from the sun (earth) ({\tt SUSPECT} RGE
code). RGE code systematic errors were estimated to be $\sim$ 0.5\%
and 0.6\% respectively. Note that for these quantities additional
systematic errors in the predictions (at least a factor 2
\cite{pgpriv}) due to uncertainties in the $\chioi$ density and
$\chioi$-$\chioi$ annihilation rates in the earth and sun are
applicable. Thus the statistical and systematic errors associated with
the ATLAS predictions and RGE codes are again effectively negligible.

\section{Discussion and Conclusions}
These results demonstrate both the effectiveness of ATLAS mSUGRA
parameter measurements for predicting SUSY Dark Matter properties and
the excellent agreement which now exists between RGE codes. For SPS1a
the statistical and systematic errors arising from these sources are
expected to be negligible compared with other sources of systematic
error in the estimates of Dark Matter properties. Of course for higher
mass scale models in e.g. the focus point, co-annihilation or
rapid-annihilation regions of mSUGRA parameter space, the smaller
number of mass constraints may well lead to considerably increased
statistical errors on mSUGRA parameters and hence the quantities
calculated here. The SPS1a benchmark model considered here should
therefore probably be considered to be something of a `best-case'
scenario. More work is clearly needed to study other more difficult
cases.

It is interesting to note that the result of the (mSUGRA)
model-dependent calculation of $\ohsq$ performed here could be assumed
to be more model-independent provided certain criteria were met by the
SUSY signal observed in ATLAS. This result should be valid for any
`bulk' region SUSY model with dominant $\chioi$ annihilation into
$\tau$ or slepton/sneutrino pairs \footnote{At SPS1a these processes
contribute $\sim$96\% of the total (co)-annihilation cross-section
according to {\tt MICROMEGAS}.} and $\staui$, $\slr$ and $\chioi$
masses and $\tan{\beta}$ compatible with the observed invariant mass
end-point positions and branching fractions. Additional measurements
would be needed in order to perform a completely-model independent
estimate of the relic density.  Examples are the $\tchi^0_3$ mass,
which would allow complete determination of the neutralino mixing
matrix (if assumed real), additional information on the $\tilde\tau$
sector, and a measurement of the masses of the heavy higgses. Work is
in progress in order to evaluate how well the presently available
measurements can constrain a generic MSSM, and to identify the crucial
measurements needed in order to achieve model-independence.

\section*{Acknowledgements}
DRT wishes to thank Genevieve Belanger and Paolo Gondolo for
discussions. DRT wishes to acknowledge PPARC for support.

\end{document}